\documentclass[a4paper,11pt]{article}

\usepackage[dvips]{graphicx}
\usepackage{amsfonts, color, float, amsmath}
\usepackage{amssymb}\usepackage{subfigure}
\usepackage{url}
\usepackage{array}
\usepackage[normalem]{ulem}

\usepackage{slashed}
\usepackage[dvipsnames]{xcolor}

\usepackage[small,bf,sf]{caption}
\usepackage{microtype}
\usepackage{cite}
\usepackage{authblk}

\usepackage[bottom]{footmisc}
\usepackage{booktabs}
\usepackage{hyperref}
\usepackage[capitalise]{cleveref}

\usepackage{xspace}

\usepackage{colortbl}
\definecolor{myblue}{rgb}{0.8,0.85,1}
\definecolor{light-gray}{gray}{0.95}

\hypersetup{colorlinks=true,
citecolor=ForestGreen,
anchorcolor=CornflowerBlue,
linkcolor=NavyBlue,
urlcolor=Mahogany,
linkbordercolor={1 1 0}}

\makeatletter
\newcommand{\thickhline}{\noalign {\ifnum 0=`}\fi \hrule height 1pt
\futurelet \reserved@a \@xhline}
\newcolumntype{"}{@{\hskip\tabcolsep\vrule width 1pt\hskip\tabcolsep}}
\makeatother

\allowdisplaybreaks

\newcommand{\MSb}{{\overline{\rm MS}}}
\newcommand{\wh}{\widehat}

\newcommand{\Lb}{large-$\beta_0$\xspace}

\renewcommand{\Im}{{\rm Im}}

\def\beq {\begin{equation}}
\def\eeq {\end{equation}}
\def\bea {\begin{eqnarray}}
\def\eea {\end{eqnarray}}

\def\nn {\nonumber}

\def \wh{\widehat}

\AtBeginDocument{}

\title{\LARGE {\bf \sffamily \boldmath Higher-order QCD corrections to $H\to b\bar b$ from rational approximants}}

\author[a,b]{Diogo Boito}
\author[a]{Cristiane Y. London\thanks{cristiane.london@usp.br (corresponding author)\vspace{0.3cm}}}
\author[c]{Pere Masjuan}

\affil[a]{\it Instituto de F\'isica de São Carlos, Universidade de São Paulo, CP 369, 13560-970, São Carlos, SP, Brazil\vspace{0.3cm}}

\affil[b]{\it University of Vienna, Faculty of Physics, Boltzmanngasse 5, A-1090 Wien, Austria\vspace{0.3cm}}

\affil[c]{\it Grup de F\'isica Te\`orica, Departament de F\'isica, Universitat Aut\`onoma de Barcelona, and Institut de F\'isica d'Altes Energies (IFAE), The Barcelona Institute of Science and Technology (BIST), Campus UAB, E-08193 Bellaterra (Barcelona), Spain }

\date{}

\begin{document}
\begin{flushright}
{\small UWThPh 2021-6 \\ \today}
\end{flushright}

\vspace*{-0.7cm}
\begingroup
\let\newpage\relax
\maketitle
\endgroup
\date{}

\vspace*{-1.0cm}
\begin{abstract}
\noindent
We use rational approximants to study missing higher orders in the massless scalar-current quark correlator. We predict the yet unknown six-loop coefficient of its imaginary part, related to $\Gamma(H\to b \bar b)$, to be $c_5=-6900\pm 1400$.
With this result, the perturbative series becomes almost insensitive to renormalization scale variations and the intrinsic QCD truncation uncertainty is tiny. This confirms the expectation that higher-order loop computations for this quantity will not be required in the foreseeable future, as the uncertainty in $\Gamma(H\to b \bar b)$ will remain largely dominated by the Standard Model parameters.
\end{abstract}

\thispagestyle{empty}

\vspace*{0.0cm}
\tableofcontents

\setcounter{page}{1}

\section{Introduction}

Without direct evidence for new physics from the Large Hadron Collider (LHC), searches for physics beyond the Standard Model will rely, in the near
future, on indirect evidence from precise measurements that could reveal small deviations from the Standard Model (SM) predictions. Higgs
physics plays a central role in this program. So far, the LHC has not found any statistically significant evidence of deviations from SM expectations in this sector, but the
uncertainties in measurements of the required cross-sections and decay rates, for example, remain rather significant. With future facilities, such as the FCC-ee
\cite{FCC:2018byv}, which is planned to run as a Higgs factory at $\sqrt{s}=250$~GeV, the experimental status should improve
dramatically. This will require a large effort on the theory side to reach the required precision for these processes~\cite{Heinemeyer:2021rgq,Freitas:2019bre}, lest the tests of the SM remain inconclusive.

The dominant decay channel for the SM Higgs boson is in a $b\bar b$ pair. At the LHC, however, due to the contamination from the hadronic background, the measurement of the decay $H\to b\bar b$ is challenging and the first results were reported only in 2018~\cite{ATLAS:2018kot,CMS:2018nsn}. The precise description of this decay in the SM requires the calculation of higher-order perturbative QCD
contributions, which are the dominant source of radiative corrections. The result is known with full quark-mass dependence at NLO~\cite{Braaten:1980yq,Sakai:1980fa} and at order $\alpha_s^2$ from Ref.~\cite{Chetyrkin:1998ix}. Beyond NNLO,
quark-mass effects are treated perturbatively, with the leading order results obtained for massless bottom quarks (in the quark propagators), receiving small $m_b/m_H$ corrections (top mass effects, on the other hand, can be sizeable). In this framework, the corrections in the limit $m_b/m_H\to 0$ are completely known up to $\alpha_s^4$ thanks to the impressive five-loop computation of the terms originating from the Higgs-bottom Yukawa coupling~\cite{Baikov:2005rw,Herzog:2017dtz} and the work of Ref.~\cite{Davies:2017xsp} where the top-quark induced terms are calculated in the effective field theory framework, with the top quark integrated out. (A complete discussion of further corrections, electroweak and mixed electroweak for example, can be found in the dedicated reviews such as Ref.~\cite{Spira:2016ztx}.)

In this work, we are concerned with estimating higher orders in the Higgs-bottom Yukawa induced contributions in the massless limit, starting at N5LO, or six loops, which may never be calculated exactly, with the aim of reassessing the
intrinsic truncation error associated with perturbative QCD.
We will employ different types of rational approximants, or Pad\'e approximants~\cite{Baker1975essentials,Baker1996pade}, in
order to exploit the exact knowledge of the first
four terms in the perturbative series to reconstruct the series to even higher orders. Variants of this method have been used for a long time, in different processes. Here, our strategy follows what was developed for hadronic $\tau$ decays~\cite{Boito:2018rwt}, where the QCD corrections,
arising from the massless Adler function, are also known up to $\mathcal{O}(\alpha_s^4)$ (N4LO)~\cite{Baikov:2008jh,Herzog:2017dtz}. Ref.~\cite{Boito:2018rwt} is the main reference of the present work.

The strategy can be
summarized as follows. First, we make use of the results for the relevant correlator in the \Lb limit of QCD~\cite{Beneke:1994qe,Broadhurst:1994se,Beneke:1998ui}, which provides an all-order
realistic series, with renormalon singularities and the associated factorial divergences, in which different approximants can be tested and their
precision can be judged by comparison with the exact results. In fact, we will exploit mainly rational approximants built to the Borel transform
of the perturbative series, since this transform suppresses the factorial divergences and can have a finite radius of convergence. This type of method is
sometimes referred to as ``Pad\'e-Borel" and benefits from the fact that rational approximants are ideal to approximate functions with isolated
poles --- here those poles are the {\it renormalons} of perturbation theory. Using the results from the \Lb limit, we are able to design
the ideal strategy to the problem at hand. This strategy is then employed in full QCD, where our knowledge is limited to the first four non-trivial coefficients of the series.

We will show that the so-called D-log Pad\'e approximants play a central role in our analysis. This type of
approximant, which is less common in the literature, is particularly suitable to deal with functions containing branch points or poles with higher multiplicity. In QCD, the renormalon poles evolve
into branch points~\cite{Beneke:1998ui,Jamin:2021qxb}, which is precisely why D-log Pad\'es are so appealing when working with the Borel transformed perturbative series.

Through a careful and systematic use of these rational approximants, we are able to obtain a model-independent prediction for the first unknown coefficient
of the scalar $q\bar{q}$ correlator, the $\alpha_s^5$ term (N5LO), with an associated uncertainty --- a benefit of our method. We also obtain
estimates for even higher orders, starting at N6LO, albeit with an increasingly large error. With these results we can calculate the total
perturbative QCD contribution to the decay width of the Higgs boson into $b\bar{b}$ with an associated uncertainty for missing higher orders.
The use of our results for the N5LO coefficient further decreases the perturbative uncertainty.
Our results confirm the expectation that the dominant QCD contributions~arising from the scalar $q\bar q$ correlator are under very good control, with a small error from the truncation of the series, and they reinforce that the major limiting factors for this decay are, and will be in the foreseeable future, the precision of the $b$-quark mass and of the strong coupling.

This paper is organized as follows. We start in Sec.~\ref{sec:pade} with a brief overview of elements of Pad\'e theory with focus on our applications. In Sec.~\ref{sec:correlatorqcd} we define the massless scalar correlator and related physical observables and present
their perturbative expansion up to the last known term, the $\mathcal{O}(\alpha_s^4)$. In Sec.~\ref{sec:largebeta0} we discuss in detail a specific application in the \Lb limit, which serves as a proof of concept.
In Sec.~\ref{sec:paqcd}, we present our results in QCD and their impact on the decay width for $H\to b\bar b$. Our conclusions are given in Sec.~\ref{sec:conclusions}.

\section{Pad\'e approximants in a nutshell}\label{sec:pade}

In this section we give an overview of the most important concepts about Pad\'e approximants and their variants that are used in the presented work. A more detailed discussion focussed on a similar application can be found in Ref.~\cite{Boito:2018rwt}; broader reviews of the topic can be found in Refs.~\cite{Baker1975essentials,Baker1996pade,MasjuanQueralt:2010hav}.

Let us consider a function $f(z)$ whose series expansion in the complex plane around $z=0$ is given by
\beq
f(z) = \sum_{n=0}^\infty f_n z^n . \label{eq:fz}
\eeq
A Pad\'e approximant to the function $f(z)$ \cite{Baker1996pade}, denoted as $P_N^M (z)$, is defined as the ratio of two polynomials $Q_M(z)$ and $R_N(z)$ of order $M$ and $N$ respectively, where the definition $R_N(0) = 1$ is employed without loss of generality
\beq
P_N^M (z) = \dfrac{Q_M(z)}{R_N(z)} = \dfrac{a_0 + a_1 z + \cdots + a_M z^M}{1 + b_1 z + \cdots + b_N z^N} . \label{eq:pa}
\eeq
The PA makes a ``contact'' of order $M+N$ with the expansion of the function around $z=0$ \cite{Masjuan:2007ay}; which means that, when $P_N^M (z)$ is expanded around this point, it will reproduce exactly the first $M+N+1$ coefficients $f_n$, thus returning values for $a_0,...,a_M$, and $b_1$,...,$b_N$ and predicting the coefficients of order higher than $M+N$.

In the Pad\'e approximant theory, Pomerenke's theorem guarantees the convergence of Pad\'es sequences $P^{N+k}_N$, for $k \geq -1$, to the original function, as long as $f(z)$ is meromorphic, either being Stieltjes or not \cite{Masjuan:2007ay,Masjuan:2008fr,Masjuan:2009wy}. Accordingly, these Pad\'es are convergent everywhere in any compact set of the complex plane, excluding a set of zero area which contains the poles of the original function \cite{Baker1996pade}. The sequence of PAs reproduces the
analytic structure of $f(z)$, its poles and residues, in a hierarchical fashion: poles that lie closer to the origin will be well reproduced already at lower orders, while poles (and the associated residues) that are further away from the origin will not correspond to any singularity of the original function, and can only be considered as effective poles. In the process, extraneous poles can be generated, but the
theorem also ensures that, in any compact region in the complex plane, these poles of the PAs will move far from this region when the order of the PA is increased or they will appear in combination with near-by zeros of the numerator $Q_M(z)$, constituting what is known as \textit{defects} or Froissart doublets. These defects play an important role in the analysis of the convergence of the series of PAs, since they entail a cancellation
which effectively reduce the order of the PA.

For meromorphic functions, some of the poles and residues of the PA may be complex, even if the original function does not have complex poles \cite{Masjuan:2007ay,Masjuan:2008fr,Masjuan:2009wy}. These poles of the PA cannot be identified with any singularity of the original function, but it remains possible to employ those Pad\'es to study the function away from the complex poles. However, in their vicinity the Pad\'e approximation deteriorates. These complex poles are transient: they appear and disappear when the order of the Pad\'e is raised \cite{Masjuan:2007ay,Masjuan:2008fr,Masjuan:2009wy}.

In the problem we are dealing with, the Borel transform in the \Lb limit contains only isolated renormalon poles.
However, the singularities of the Borel transform of the perturbative series in full QCD become superimposed cuts instead of isolated poles. Although there are no convergence theorems in Pad\'e theory for generic functions with cuts (except for Stieltjes functions) experience shows that
PAs can approximate functions with cuts remarkably well. This is achieved by an accumulation of poles along the cut, which mimics the singularity structure of the original function~\cite{Baker1996pade,MasjuanQueralt:2010hav}.

A different kind of PAs that is very useful for functions with branch points or poles with higher multiplicity are the D-log Pad\'e approximants \cite{Baker1996pade,Boito:2018rwt}. Let us consider the function
\beq
f(z) = A(z) \, \dfrac{1}{(\mu - z)^\gamma} + B(z) ,
\eeq
where $A(z)$ and $B(z)$ are functions with little structure and analytic at $z=\mu$. We are primarily interested in the case in which $f(z)$ has a branch point at $z=\mu$, and accordingly $\gamma$ is a non-integer number, but in reality this condition is not essential. We can define now a new function $F(z)$ near $z=\mu$ as \cite{Baker1996pade}
\beq
F(z) = \dfrac{\mathrm{d}}{\mathrm{d} z} \ln{f(z)} \approx \dfrac{\gamma}{(\mu - z)} . \label{eq:Fdlog}
\eeq
The function $F(z)$ has a simple pole and the residue is the exponent of the cut of $f(z)$. Thus, with $\bar P_N^M(z)$ being the Pad\'e constructed to $F(z)$ defined above, the D-log Pad\'e, $\mathrm{Dlog}_N^M(z)$, of $f(z)$ is given by the expression below \cite{Baker1996pade,Boito:2018rwt}
\beq
\mathrm{Dlog}_N^M(z) = f(0) \exp{\left[\int \mathrm{d}z \, \bar{P}_N^M(z) \right]} . \label{eq:dlogpa}
\eeq
Because of the derivative in \cref{eq:Fdlog}, the $f(0)$ term is lost and must be reintroduced in order to correctly normalize the D-log Pad\'e. The $\mathrm{Dlog}_N^M$ then reproduces exactly the first $M+N+2$ coefficients of $f(z)$ (one order more than the usual $P_N^M$) and can be used to predict the $(M+N+3)$-th coefficient and higher. The D-log Pad\'e is no longer a rational approximant, however the function $F(z)$ is meromorphic and, because of that, it is easily approximated by the Pad\'e $\bar P_N^M$.

In principle, this type of approximant offers a way to determine the branch point and the exponent of the cut of the original function $f(z)$ from the study of the PA to $F(z)$ around its pole. Since no assumption about $\mu$ or $\gamma$ is made, their estimates are exclusively obtained from the series coefficients.

\section{Scalar correlator}\label{sec:correlatorqcd}
\label{sec:ScalarCorr}

As already mentioned, the decay width of the Higgs boson into the pair $b\bar{b}$ is known up to fourth order in QCD \cite{Baikov:2005rw,Herzog:2017dtz} in the massless limit and is related to the imaginary part of the quark-antiquark scalar current correlator. This correlator, that is not directly associated with any physical quantity, is defined as
\beq
\Pi (p^2) \equiv i \int \mathrm{d}x \, \mathrm{e}^{ipx} \, \langle \Omega | T\{j(x)j^\dag(0)\} | \Omega \rangle , \label{eq:defscalarcorrelator}
\eeq
with $|\Omega \rangle$ representing the non-perturbative QCD vacuum. The scalar current $j(x)$ arises from the interaction between the Higgs and the bottom quarks and is given by
\beq
j(x) = m_q\, \overline{q}_f(x) q_f(x) , \label{eq:scalarcurrent}
\eeq
where $q(x)$ is the quark field treated as massless in loop calculations. The mass of the quark, $m_q$, is introduced to ensure renormalization group invariance for the physical observables.

The purely perturbative expansion of $\Pi(s)$ in powers of the strong coupling is given by the following general form \cite{Jamin:2016ihy}
\beq
\Pi(s) = - \dfrac{N_c}{8\pi^2} \, s\, m_q^2(\mu) \sum_{n=0}^\infty a_s^n(\mu) \sum_{k=0}^{n+1} d_{n,k} L^k , \label{eq:piqcd}
\eeq
where $s \equiv p^2$. The quark mass $m_q$ and the QCD coupling $a_s\equiv\alpha_s/\pi$, both in the $\MSb$ scheme, are renormalized at the scale $\mu$,\footnote{We will often omit the dependence of $a_s$ and $m_q$ on the renormalization scale.} which appears in the logarithms $L \equiv \ln{(-s/\mu^2)}$ as well. Setting the scale $\mu^2=-s$ the logarithms are resummed.

The coefficients $d_{n,0}$ depend on the conventions related to the renormalization procedure and do not contribute in any physical quantity. Furthermore, the coefficients $d_{n,k}$, with $k>1$, can be obtained from the RGE of the scalar correlator~\cite{Chetyrkin:1996sr}.
Hence the only independent coefficients for each perturbative order are $d_{n,1}$, which are known analytically up to fourth order and whose numerical values for five active flavours, $N_f=5$, and $N_c=3$ are \cite{Baikov:2005rw,Herzog:2017dtz}
\beq
d_{1,1} = 5.6667 , \qquad d_{2,1} = 42.032 , \qquad d_{3,1} = 353.229 , \qquad d_{4,1} = 3512.2 . \label{eq:dn1qcd}
\eeq

The physical decay width for $H\to b\bar b$ is related to the imaginary part of the scalar correlator $\Pi(s)$ as
\beq
\Gamma \left( H \rightarrow b\bar{b} \right) = \dfrac{1}{m_H\, v^2} \,\operatorname{Im} \Pi(s) ,
\eeq
where $v^2$ is the Higgs vacuum expectation value.
Setting the renormalization scale to $\mu^2 = s \equiv m_H^2$, so that the logarithms are summed, the general perturbative expansion for the imaginary part of $\Pi(s)$ is
\beq
\operatorname{Im} \Pi (s) = \dfrac{N_c}{8\pi} \, m_q^2(m_H) \, s \sum_{n=0}^\infty a_s^n(m_H) \sum_{k=0}^{[n/2]} d_{n,2k+1} (i\pi)^{2k} , \label{eq:impiqcd}
\eeq
with $[x]$ representing the integer part of $x$. For $N_f=5$, the expansion which is known up to fourth order in QCD is
\begin{align}
&\operatorname{Im} \Pi (s) = \dfrac{N_c}{8\pi} \, m^2_q(m_H) \, s\left[ 1+ \sum_{n=1}^\infty c_n a_s^n \right] = \dfrac{N_c}{8\pi} \, m^2_q(m_H) \, s\left[ 1+ F(a_s) \right]\nn \\
&= \dfrac{N_c}{8\pi} \, m^2_q(m_H) \, s \left[ 1 + 5.6667 \, a_s + 29.1467 \, a_s^2 + 41.7576 \, a_s^3 - 825.747 \, a_s^4 + \dots \right]. \label{eq:impiqcdvalor}
\end{align}
In passing, we defined the function $F(a_s)$ which contains the $\alpha_s$ corrections to $\operatorname{Im} \Pi (s)$ and is referred to as the reduced imaginary part.
As seen above, a negative coefficient appears in the term of order $a_s^4$ of the imaginary part of $\Pi(s)$. Since the imaginary part of the scalar correlator is a physical observable, it satisfies a homogeneous RGE and the logarithms can be reinstated from the expressions above.

Another renormalization group invariant quantity is the second derivative of the scalar correlator with respect to $s$~\cite{Jamin:2016ihy,Jamin:2021qxb}, which removes the two scheme-dependent subtraction constants. Its general perturbative expansion can be easily obtained from Eq.~(\ref{eq:piqcd}) and, resumming the logarithms with the scale $\mu^2=-s$, we have~\cite{Jamin:2016ihy}
\beq
\Pi''(s)
= - \dfrac{N_c}{8\pi^2} \dfrac{m^2_q(m_H)}{s} \left[ 1 + \sum_{n=1}^\infty r_n a_s^n \right] , \label{eq:pi''qcd}
\eeq
where $r_n=\left( d_{n,1} + 2 d_{n,2} \right)$ and whose numerical coefficients for $N_f=5$ are
\beq
\Pi''(s) = - \dfrac{N_c}{8\pi^2} \dfrac{m^2_q(m_H)}{s} \left[ 1 + 3.6667 \, a_s + 12.8098 \, a_s^2 + 39.6839 \, a_s^3 + 153.955 \, a_s^4 + \dots \right] . \label{eq:pi''qcdvalue}
\eeq

Since all the series discussed here are (at best) asymptotic, it is convenient to work with their Borel transform, which
suppresses the factorial divergence of the series coefficients. Without loss of generality, we always consider the Borel transform of functions starting at $\mathcal{O}(\alpha_s)$, such as $F(a_s)$ in Eq.~(\ref{eq:impiqcdvalor}). We refer to these functions as ``reduced functions". Let
\beq
R(\alpha_s) = \sum_{k=0}^\infty r_k \, \alpha_s^{k+1} ,
\eeq
be an asymptotic expansion of a given observable $R$. We define its Borel transform as
\beq
B[R](u) = \sum_{k=0}^\infty \dfrac{r_k}{k!} \left(\frac{2\pi}{\beta_1}\right)^{k+1}\, u^k , \label{eq:borel}
\eeq
where $\beta_1$ is the first coefficient of the QCD $\beta$ function (see App.~\ref{app:betagamma}).
The Borel integral, which gives the value of $R$ summed in the Borel sense (provided the integral exists), is
\beq
R = \int\displaylimits_0^\infty {\rm d} u\, \mathrm{e}^{-2\pi u/(\beta_1\alpha_s)} B[R](u) .\label{eq:intborel}
\eeq

The singularities of the Borel transform in the complex $u$ plane, which in our case are the renormalons of perturbation theory, govern the behavior of the perturbative series at intermediate and higher orders \cite{Beneke:1998ui}. Poles on the negative real axis generate sign alternating coefficients. These poles are of UV origin in the applications we discuss in this paper. Poles of IR origin are located on the positive real axis and obstruct the integration in Eq.~(\ref{eq:intborel}). A prescription to circumvent these poles must be adopted, which generates an imaginary ambiguity in the Borel sum of the series. It is expected on general grounds that these ambiguities cancel against the non-perturbative corrections from higher-dimensional OPE condensates. In the presence of several renormalon singularities, the one closest to the origin dominates the series behavior at intermediate and large orders.

In this work, in the majority of cases, we will build PAs to the Borel transforms of the series in $\alpha_s$. This procedure, sometimes referred to as Pad\'e-Borel approximants, is known empirically to lead to faster convergence~\cite{Samuel:1995jc} for asymptotic series of the type we have here~(see Ref.~\cite{Costin:2021bay} and references therein).

\section{\boldmath A proof of concept in the \texorpdfstring{large-$\beta_0$}{large-beta0} limit}\label{sec:largebeta0}

The aim of this section is to employ our technique in a realistic case, as a proof of concept of our method. We choose to highlight an application of D-log Pad\'e approximants, since these are very seldom used in the literature. We apply them to the Borel transform of $\Pi''(s)$ in the \Lb limit of QCD~\cite{Beneke:1994qe,Broadhurst:1994se}.
We choose to work initially in this limit because it provides an entirely consistent simplified model where higher-order corrections are known to all orders in the coupling. The series exhibit the renormalon divergences, although they appear as simple or double poles (and not branch cuts as in full QCD)~\cite{Beneke:1998ui}. The large-$\beta_0$ result is obtained from the leading $N_f$ terms, which can be calculated for the scalar
correlator by considering light-quark bubble loop corrections to the gluon propagators in the 2-loop result. Then, through the procedure
known as naive non-abelianization~\cite{Broadhurst:1994se,Beneke:1994qe}, in which the fermionic contribution to the QCD beta function is replaced by the full one-loop beta function
coefficient ($\beta_1$ in our notation), a set of non-abelian terms is effectively introduced, generating a realistic result to all orders in $\alpha_s$.

The massless scalar correlator in the large-$\beta_0$ limit was first calculated by Broadhurst, Kataev and Maxwell~\cite{Broadhurst:2000yc}, who obtained a result for its Borel transform in closed form. With this result and the recent discussion in Ref.~\cite{Jamin:2016ihy}
one finds that the Borel transform of $\operatorname{Im}\Pi(s)$, in terms of the renormalization group invariant (RGI) quark mass (see App.~\ref{app:betagamma}), can be written as
\beq
B [\wh{F}^C_{L\beta}] (u) = \dfrac{3C_F}{\beta_1} \left[ \dfrac{\sin{\pi u}}{\pi u} \left[ 1 + u \, G_D(u) \right] \mathrm{e}^{(C + 5/3) u} - 1 \right] \dfrac{1}{u} , \label{eq:borelfsrgi}
\eeq
where $C_F=4/3$ and $C$ is a parameter that controls the renormalization scheme, with $C=0$ corresponding to $\MSb$, our preferred choice.
The function $G_D(u)$ can be written as
\beq
G_D(u) = \dfrac{2}{1-u} - \dfrac{1}{2-u} + \dfrac{2}{3} \sum_{k=3}^\infty \dfrac{(-1)^k}{(k-u)^2} -\dfrac{2}{3} \sum_{k=1}^\infty \dfrac{(-1)^k}{(k+u)^2} . \label{eq:gd}
\eeq
Due to the zeros of the function $\frac{\sin{\pi u}}{\pi u}$, the Borel transform (\ref{eq:borelfsrgi}) has no renormalons in $u = 1$ and $u = 2$ and all poles are simple poles. (Note that the expression is regular at $u=0$.)

The Borel transform of the reduced second derivative of $\Pi(s)$ written in terms of the RGI quark mass in large-$\beta_0$ is
\beq
B [\wh{D}_{L\beta}^C] (u) = \dfrac{3C_F}{\beta_1} \left[ (1 - u) \left[ 1 + u \, G_D(u) \right] \mathrm{e}^{(C + 5/3) u} - 1 \right] \dfrac{1}{u} . \label{eq:boreldsrgi}
\eeq
This Borel transform has a simple IR pole at $u=2$, associated with the gluon condensate corrections. The other renormalon poles, at $u=-1,-2,-3,\dots$ and $u=3,4,5,\dots$, are all double poles. Since the second derivative of the scalar correlator is physical, there is no renormalon at $u=1$.
At sufficiently high orders the perturbative series coefficients will be dominated by the UV pole at $u=-1$, which is the pole closest to the origin, and will be sign alternating.

Comparing the Borel transforms of $\operatorname{Im} \Pi(s)$ and $\Pi''(s)$ one sees that in the first the double poles are reduced to simple poles, which should improve the performance of the PAs. However, this is done at a price: the PA will need to reproduce the function $\sin(\pi u)$ as well. In this scenario, it is advantageous to work with the Borel transform of $\Pi''(s)$ but using D-log Pad\'e approximants, which achieve the reduction of the double poles to simple ones without the complications of the prefactor of Eq.~(\ref{eq:borelfsrgi}). We focus on this case.

Reconstructing the perturbative expansion of $\wh{D}_{L\beta}^C$ in the $\MSb$ scheme with $N_f=5$ we find
\beq
\wh{D}_{L\beta} = 5.3333 \, a_s + 6.0111 \, a_s^2 + 42.8911 \, a_s^3 + 0.7729 \, a_s^4 + 1512.48 \, a_s^5 - 8410.57 \, a_s^6 + \dots . \label{eq:pi''coefrgi}
\eeq
The coefficients of Eq.~(\ref{eq:pi''coefrgi}) should not be directly compared with those of Eq.~(\ref{eq:pi''qcdvalue}), since here we are expressing the result in terms of $\widehat m_q$.\footnote{This choice is motivated by the fact that in \Lb the $\gamma_m$ function is known to all orders and the $\beta$ function is truncated at $\beta_1$. Furthermore, using $\widehat m_q$, the Borel transform assumes a much simpler form.}
The expected sign alternation of the coefficients, dictated by the UV renormalon, sets in from the sixth-order coefficient onwards.

\subsection{D-log Pad\'e approximants}

One could build standard PAs to the Borel transform of Eq.~(\ref{eq:boreldsrgi}). Although these approximants do display convergence, they are not optimal and the convergence of the procedure can be accelerated with the use of D-log Pad\'e approximants.
One of the reasons for the slower convergence of the PAs in the case of the Borel transform of $\Pi''(s)$ is the presence of an infinite number of double poles. Ideally, the approximants need to reproduce the pole multiplicity in order to achieve a good description of the series, which requires a larger number of input coefficients.
A method that can be more effective in the presence of multiple poles are the D-log Pad\'es defined in Sec.~\ref{sec:pade}. We now turn to their application in approximating the Borel transform of the series $\wh{D}_{L\beta}$, given in \cref{eq:boreldsrgi}.

We remind that the D-log Pad\'e $\mathrm{Dlog}_N^M$ is built from a PA $\bar P_N^M$ to the function $F(z)$ of Eq.~(\ref{eq:Fdlog}).
It reproduces the first $N+M+2$ coefficients and can be used to predict the $(M+N+3)$-th coefficient and higher. We will consider $\mathrm{Dlog}_N^M(u)$ with $N+M=2$ or higher, since otherwise the PAs $\bar P_N^M$ that are required contain too little information (one or two coefficients only). The perturbative coefficients predicted by the D-log Pad\'es will be denoted $\widehat r_k^P$ and are the counterpart of the exact coefficients of Eq.~(\ref{eq:pi''coefrgi}), denoted by $\widehat r_n$. We assess the convergence of the D-log Pad\'e sequence through the relative error of the predicted coefficients, defined as
\beq
\sigma_{\mathrm{rel}} = \left\lvert \dfrac{\widehat r^P_n - \widehat r_n}{\widehat r_n} \right\rvert . \label{eq:relativeerror2}
\eeq

Let us begin examining a simple example, $\mathrm{Dlog}_1^1(u)$, whose expression is
\beq
\mathrm{Dlog}_1^1(u) = \dfrac{64}{23} \dfrac{\mathrm{e}^{0.1504 + 2.5667 u}}{(1.0734 + u)^{2.1238}} . \label{eq:dlog11}
\eeq
After expanding it, the first predicted coefficient is $\wh{r}_5^P = 1127$ with an error of just 25\%.
The subsequent coefficients $\wh{r}_6^P$ and $\wh{r}_7^P$ are reproduced within 20\% and 38\%, respectively. This D-log Pad\'e approximant also predicts the sign-alternating behavior and it reproduces very precisely the leading renormalon of the Borel transform, the UV double pole at $u=-1$. Analyzing the next approximant in this sequence, $\mathrm{Dlog}_2^2$, the first predicted coefficient, $\wh{r}_7^P = 189\,406$, is also in very good agreement with the exact one: it is off by only 16\%. This D-log Pad\'e also has a pole close to the leading UV pole, at $u=-0.83$, but its multiplicity, 1.10, is less well reproduced, about half of the real value.

\begin{figure}[!t]
\begin{center}
\includegraphics[width=0.49\textwidth]{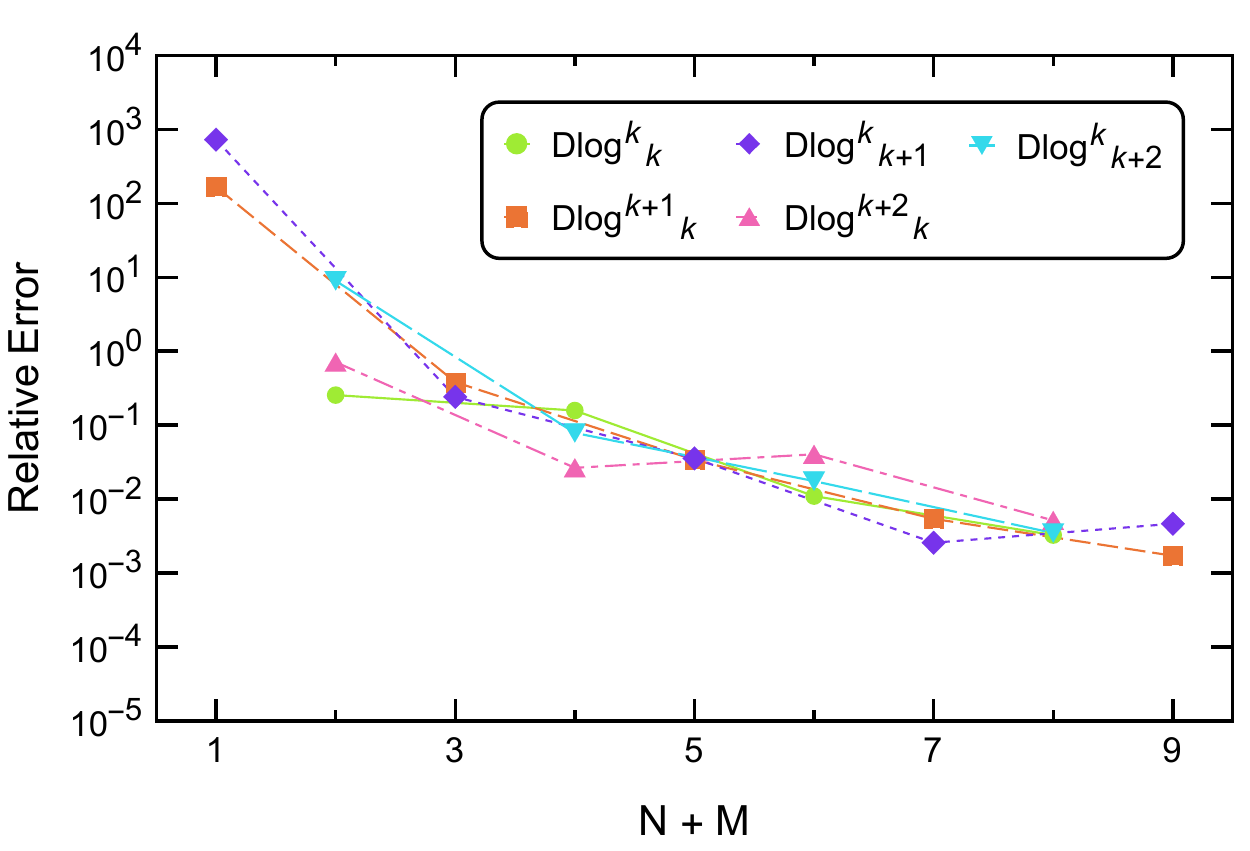}
\caption{Relative error of the first predicted coefficient $\wh{r}_n$ of $\Pi''(s)$ in the \Lb limit obtained from D-log Pad\'e approximants applied to the Borel transform of $\Pi''(s)$. Results for ${\rm Dlog}^M_N$ shown as a function of $N+M$.}\label{fig:errorelativod2pidlogrgi}
\end{center}
\end{figure}

Fig.~\ref{fig:errorelativod2pidlogrgi} shows the relative error of the first coefficient predicted by D-log Pad\'e approximants belonging to different sequences and illustrates the convergence of the procedure. One can see that the relative errors decrease when the order of the D-log is raised.
In the few cases where the error increases for higher-order approximants this can be understood in terms of the appearance of poles very far from the origin or defects (as discussed in Sec.~\ref{sec:pade}) which effectively reduce the order of the approximant.

Since our goal is to apply the procedure to QCD, it is important to focus on the D-log Pad\'es built from the first four coefficients of the series, such as the approximant of Eq.~(\ref{eq:dlog11}), which predict the coefficient of $\mathcal{O}\left(\alpha_s^5\right)$. In Fig.~\ref{fig:errorelativod2pidlogrgi}, we see that for $N+M=2$ the $\mathrm{Dlog}_2^0$ has a relative error approximately two orders of magnitude larger than $\mathrm{Dlog}_0^2$ and $\mathrm{Dlog}_1^1$. The reason for the anomalous behavior can be traced to the standard Pad\'e $\bar{P}_2^0$ used to build the $\mathrm{Dlog}^0_2$ (see \cref{eq:Fdlog,eq:dlogpa}). The approximant $\bar{P}_2^0$ has a pair of complex poles at $u= 0.1230 \pm 0.2517 i$,
not too far from the origin. As we discussed in Sec.~\ref{sec:pade}, the complex poles can appear when the function to be approximated is meromorphic but not Stieltjes, however the Pad\'e approximation to the function by definition breaks down close to these poles.
Since we are interested in the behavior around the origin, the poles are not sufficiently far and, thus, the estimates of $\mathrm{Dlog}_2^0$ can be disconsidered since $\bar{P}_2^0$ is not a reliable approximation for the function $F(z)$ defined in \cref{eq:Fdlog} in the range up to the first renormalon poles.

Finally, estimates of the Borel sum obtained from the D-log Pad\'es are very close to the true value and get better with larger $k$. Furthermore, we observe that when the order of the D-log Pad\'e is increased, the sign-alternating behavior and the dominant pole $u=-1$ are well replicated, although for the latter often with a multiplicity that is not so close to the exact value.
The quality of the results obtained from the D-log approximants in \Lb can be understood from the fact that the procedure
reduces the double poles of the Borel transform to simple poles. This type of singularity softening also happens when the D-logs are applied to functions with branch cuts, which make their use appealing in full QCD.

\subsection[Independent coefficients \texorpdfstring{$d_{n,1}$}{dn1} and partial conclusions]{Independent coefficients \boldmath \texorpdfstring{$d_{n,1}$}{dn1} and partial conclusions}\label{sec:dn1largebeta0}

As we discussed in Sec.~\ref{sec:correlatorqcd}, the only independent coefficients of the scalar correlator are the $d_{n,1}$ of \cref{eq:dn1qcd}. It is, of course, possible to determine them from the coefficients of the second derivative of $\Pi(s)$. The relation between $d_{n,1}$ and $\wh r_n$, the coefficients of $\Pi''(s)$ written in terms of the RGI quark mass, reads
\beq
d_{n,1} = \wh{r}_n - \dfrac{2}{\beta_1} \, \gamma_m^{(n+1)} + \dfrac{(n-1)}{2} \, \beta_{1} d_{n-1,1} .
\eeq
In Tab.~\ref{tab:dn1largebeta0} we show the coefficients $d_{n,1}$ obtained from the D-log Pad\'e approximants that predict the coefficient of $\mathcal{O}(\alpha_s^5)$. (The results from $\mathrm{Dlog}_2^0$ are not shown for the reasons already discussed.)

The predictions for the coefficients $d_{n,1}$ are in general very good. It should also be observed that the spread in their values, which is a measure of the associated error, is smaller than in the coefficients $\wh{r}_n$. This occurs because the coefficients $d_{n,1}$, besides depending on the predicted $\wh{r}_n$, also depend on the previous $d_{n-1,1}$ coefficients, on the $\gamma_m$-function coefficients, and $\beta_1$. The $\gamma_m$-function coefficients are exactly known to all orders in the \Lb limit while the coefficients $d_{n-1,1}$ are exactly known for all coefficients used as input in the construction of the approximants. Hence, the dispersion in the values of $d_{5,1}$, for example, arises solely from $\wh r_n$, which represents just a fraction of $d_{n,1}$.

\begin{table}[!t]
\centering
\caption{Coefficients $d_{n,1}$ in the $\MSb$ scheme from D-log Pad\'es built to the Borel transform of $\Pi''(s)$ together with their exact values in the large-$\beta_0$ limit highlighted in blue.}
\begin{tabular}{cllll}
\thickhline
& $d_{5,1}$ & $d_{6,1}$ & $d_{7,1}$ & $d_{8,1}$ \\ \hline
\rowcolor[RGB]{242,255,255}
Large-$\beta_0$ (exact) &6798 & 56\,756 & 816\,323 & $8.86 \times 10^6$ \\

$\mathrm{Dlog}_0^2$ & 5722 & 49\,141 & 573\,274 & $7.31 \times 10^6$ \\
$\mathrm{Dlog}_1^1$ & 6412 & 54\,721 & 731\,481 & $8.48 \times 10^6$ \\\thickhline
\end{tabular} \label{tab:dn1largebeta0}
\end{table}

The explorations of this section exemplify the use of the D-log approximants in a realistic case. The construction of standard Pad\'e approximants can, of course, also be explored in the \Lb limit, and they also display convergence, although somewhat slower than in the case of the D-log Pad\'es to $\Pi''(s)$. An important point is that we were able to identify the reason behind the bad predictions of some pathological approximants, that must be discarded. This is standard procedure; PAs must always be applied judiciously and critically \cite{Masjuan:2007ay,Boito:2018rwt}. Our explorations in \Lb favour approximants built to the Borel transform of the physical quantities that derive from the scalar correlator, namely $\operatorname{Im} \Pi(s)$ and $\Pi''(s)$, and in particular the use of the D-log Pad\'e approximants to $\Pi''(s)$ which proved to be superior than other alternatives.

In general, however, we observe that the results for the scalar correlator are less good than the results obtained with the same procedure applied to the massless QCD Adler function~\cite{Boito:2018rwt}. Approximants that predict the coefficient of order four were not particularly good in the scalar case; it is essential to have four coefficients at least to make meaningful predictions. The reason for that is linked to more complicated structure of the Borel transforms in the case of the scalar correlator, which reflect the non-vanishing anomalous dimension.

Other methods to accelerate convergence, such as continuous scheme changes, departing from the $\MSb$ value $C=0$, can be exploited in the \Lb limit~\cite{Jamin:2016ihy,Boito:2018rwt,Boito:2016yom}. But since the application of these methods is not so straightforward in full QCD, they become somewhat academical. We will not pursue their investigation here and refer to Ref.~\cite{Boito:2018rwt} for analogous explorations in a similar context.

\section{Results in QCD}\label{sec:paqcd}

In the previous section, we discussed an application of our method to results in the \Lb limit.
The D-log Pad\'e approximants to the Borel transform of the second derivative of $\Pi(s)$ proved to be most efficient with the number of coefficients we have in QCD, for the reasons already discussed. In QCD, the renormalons of the perturbative series are at the same position of the renormalons in the large-$\beta_0$ limit, but they become branch cuts instead of isolated poles \cite{Beneke:1998ui,Jamin:2021qxb}. Even though there are no theorems that state the convergence of the Pad\'e approximants to series with branch cuts in general, in many practical applications there are indications that this convergence happens and the mechanism for this apparent convergence is understood~\cite{Costin:2021bay}. The use of D-log Pad\'e approximants remains appealing, since these approximants are designed to deal with functions that have branch cuts.

In this section we will obtain the higher-order QCD corrections to the decay rate of the Higgs into bottom quarks. We analyze the Pad\'e and the D-log Pad\'e approximants to the Borel transform of the reduced imaginary part and of the second derivative of the scalar correlator in QCD.

\subsection[Approximants to \texorpdfstring{Im$\Pi(s)$}{Im Pi(s)} in QCD]{Approximants to \boldmath \texorpdfstring{Im$\Pi(s)$}{Im Pi(s)} in QCD}

We apply now the Pad\'e-Borel method to the Borel transform of the perturbative series of the reduced imaginary
part of $\Pi(s)$ as a function of the scale-dependent quark mass. The perturbative coefficients of this series in
QCD are given in \cref{eq:impiqcdvalor} and the Borel transform is obtained with Eq.~(\ref{eq:borel}).
The branch-cut singularities of the Borel transform are
expected to be at $u=-1,-2,\dots$ and $u=2,3,4,\dots$.

The application of standard PAs to this Borel transform leads to very discrepant results, with different signs and orders of magnitude for the coefficients of order 5 and 6. An exploration in the \Lb limit supports this finding, and the PAs are also not ideal in that limit. The difficulties of the PAs can be understood from the general structure of the Borel transform of Eq.~(\ref{eq:borelfsrgi}). First, in QCD the prefactor $\sin(\pi u)$ is unknown, but in similar applications to the Adler function it has been shown that the same factor is present in QCD~\cite{Brown:1992pk,Boito:2020hvu} provided a change to the so-called $C$ scheme is used~\cite{Boito:2016yom} --- we expect the same to happen here. However, the poles turn into branch points which means that we should not expect exact cancellations from this prefactor. This means that applications of our procedure to $\operatorname{Im} \Pi(s)$ are not favoured, since the branch cuts remain branch cuts.
The instabilities we find and the general structure of the Borel transform inferred from the \Lb results lead us to conclude that PAs are not ideal in this case and these results should be discarded.

We turn now to D-log Pad\'e approximants, which are arguably superior in this case, since we are dealing with a
function that has superimposed branch cuts. The approximants that forecast the last known coefficient, $c_4$, are
$\mathrm{Dlog}_1^0$ and $\mathrm{Dlog}_0^1$. The prediction of the first one is not particularly good, the relative
error is 88\%, but the estimate of the second one is quite close to the original value, it has an error of mere 9\%.
The results for the D-log Pad\'es that predict the first unknown coefficient of $\operatorname{Im}\Pi(s)$,
$c_5$, are in the second and third rows of Tab.~\ref{tab:coefimpi}, except for $\mathrm{Dlog}_2^0$ because the Pad\'e $\bar{P}_2^0$ used to build this
approximant has a pair of complex poles close to the origin. We can notice that the coefficients predicted in
Tab.~\ref{tab:coefimpi} are similar and very stable. The results from these two D-log approximants are the most reliable in this case and will be part of our final results. Finally, we observe that the predicted coefficients in the second and third lines of Tab.~\ref{tab:coefimpi} do not show a systematic sign alternation, which may indicate that in QCD the dominance of the leading UV singularity is postponed to higher orders (as observed in other contexts~\cite{Boito:2018rwt,Boito:2021wbj}).

\begin{table}[!t]
\centering
\caption{Perturbative coefficients of $\operatorname{Im} \Pi(s)$, $c_n$, for $N_f=5$ predicted by PAs and D-log PAs in QCD. In ``method" we indicate if the results were obtained from the Borel transforms or from the $\alpha_s$ expansion.}
\begin{tabular}{clccccc}
\thickhline
& method & $c_5$ & $c_6$ & $c_7$ & $c_8$ & $c_9$ \\ \hline
$\mathrm{Dlog}_1^1$ & $B[\operatorname{Im} \Pi(s)]$ & $-5598$ & 36\,053 & 611\,562 & $-1.84 \times 10^6$ & $-7.95 \times 10^7$ \\
$\mathrm{Dlog}_0^2$ & $B[\operatorname{Im} \Pi(s)]$ & $-5526$ & 37\,812 & 616\,726 & $-2.21 \times 10^6$ & $-8.37 \times 10^7$ \\

$P_1^2$ & $B[\Pi''(s)]$ & $-8142$ & $-26\,272$ & 171\,244 & $2.52 \times 10^6$ & $1.09 \times 10^7$ \\
$P_2^1$ & $B[\Pi''(s)]$ & $-8198$ & $-27\,773$ & 153\,734 & $2.43 \times 10^6$ & $1.11 \times 10^7$ \\
$\mathrm{Dlog}_2^0$ & $B[\Pi''(s)]$ & $-8149$ & $-26\,626$ & 164\,374 & $2.46 \times 10^6$ & $1.06 \times 10^7$ \\
$\mathrm{Dlog}_0^2$ & $B[\Pi''(s)]$ & $-8033$ & $-24\,558$ & 163\,186 & $2.09 \times 10^6$ & $6.98 \times 10^6$ \\

$P_1^3$ & $\Pi''(\alpha_s)$ & $-8341$ & $-31\,410$ & 118\,321 & $2.41 \times 10^6$ & $1.45 \times 10^7$ \\
$P_3^1$ & $\Pi''(\alpha_s)$ & $-8340$ & $-31\,480$ & 116\,501 & $2.41 \times 10^6$ & $1.47 \times 10^7$ \\ \thickhline
\end{tabular} \label{tab:coefimpi}
\end{table}

\subsection[Results for \texorpdfstring{$\Pi''(s)$}{Pi''(s)} in QCD]{Results for \boldmath \texorpdfstring{$\Pi''(s)$}{Pi''(s)} in QCD}\label{sec:2derqcd}

We turn now to the approximants built to the Borel transform of $\Pi''(s)$ which were the basis for the optimal strategy in \Lb, described in Sec.~\ref{sec:largebeta0}. We expect the D-log Pad\'es to be efficient here as well since they can deal with the branch cuts more easily. An advantage of working with the Borel transform of $\Pi''(s)$ is that the Pad\'es do not have to reproduce the QCD counterpart of the $\sin(\pi u)$ which appears in the Borel transform of $\operatorname{Im} \Pi(s)$.

We built all PAs and D-log Pad\'e approximants that post-dict the fourth- or predict the fifth-order coefficient. The PAs $P_2^0$ and $P_3^0$ have a pair of complex poles relatively close to the origin and are discarded as per the explanations of Secs.~\ref{sec:pade} and~\ref{sec:largebeta0}. Also, we do not consider the results of $\mathrm{Dlog}_1^1$ because the Pad\'e $\bar P_1^1$ used to build this D-log has an almost defect: a pole at $u=-0.1998$ and a close-by zero at $u=-0.2170$, which, as we saw in Sec.~\ref{sec:largebeta0}, effectively reduces the order of the PA and produces untrustworthy estimates.

The post-diction of $P_1^1$ for the fourth-order coefficient of $\Pi''(s)$ is accurate: the error is about 20\%. The estimates of $r_4$ from $\mathrm{Dlog}_0^1$ and $\mathrm{Dlog}_1^0$ are also close to the exact value with an error of only $\sim 30$\%. The good quality of these results is certainly reassuring but we observe that for higher orders the predictions of these approximants can differ significantly --- a fact that is in line with our conclusion that with less than four input coefficients the quality of the predictions from the approximants deteriorates quickly.

The results for the approximants that predict the fifth order (and higher) coefficients are shown from lines three to six of Tab.~\ref{tab:coefimpi}. The results from the PAs and D-log Pad\'e approximants that pass all reliability tests are all stable and are mutually consistent. These results will also enter our final estimate for the higher order coefficients.

Regarding the renormalons, we observe that all predicted coefficients of $\Pi''(s)$ up to order 9 are positive, with no sign of the dominance of the UV renormalon. Furthermore, all the PAs that use all known coefficients have singularities on the positive real axis. $\mathrm{Dlog}_2^0$ predicts a cut on the positive real axis at $u=1.642$ with multiplicity $\gamma = 1.2799$.
All of this indicates that in QCD $\Pi''(s)$ is more dominated by the IR renormalons at intermediate orders.

\subsubsection[Pad\'es to the Series in \texorpdfstring{$\alpha_s$}{alphas}]{Pad\'es to the Series \boldmath in \texorpdfstring{$\alpha_s$}{alphas} in QCD}

A possible way to corroborate the results we found previously is to perform PAs directly to the series in powers of $a_s$. Even though these approximants are less interesting since we lose part of the connection with renormalons and experience shows that for divergent series it is advantageous to work with the Pad\'e-Borel method, they provide an additional check of the robustness of the results. We have built PAs to the series expansion in powers of $a_s$ of $\operatorname{Im}\Pi(s)$ and $\Pi''(s)$. The PAs to $\operatorname{Im}\Pi(s)$ are problematic because they have Froissart doublets or complex poles dangerously close to the origin. In addition, one can notice from \cref{eq:pi''qcdvalue} that the perturbative series of the second derivative in QCD is very regular until fourth order, i.e., there is no change of sign and the known coefficients are stable (the divergent behavior is not evident up to fourth-order). Because of these two facts, we report the results of PAs built to the expansion of $\Pi''(s)$ in powers of $a_s$.

Regarding the post-diction of the last known coefficient, the results for $r_4$ from the Pad\'es $P_1^2$ and $P_2^1$ are in good agreement with the exact value, with an error of approximately 20\%.
Results from the PAs that predict the first unknown coefficient are shown in the last two rows of Tab.~\ref{tab:coefimpi}, where we can see that they are stable.
(The results for $P_2^2$ are not on the table because it has a defect.)
Analyzing the predicted coefficients of $\Pi''(s)$ from these PAs, we can notice that all the coefficients are again positive, which corroborates the dominance of the IR renormalons at lower and intermediate orders. However, the central values of the coefficients up to seventh order given in the last two rows of Tab.~\ref{tab:coefimpi} are lower than the ones obtained before. Even though there is reason to believe the results from the Pad\'e-Borel approximants to be superior we will also use these latter results in our final values to remain fully conservative.

\subsection[{Final results and uncertainties in \texorpdfstring{$H\to b\bar b $}{H to bb}}]{Final results and uncertainties in \boldmath \texorpdfstring{$H\to b\bar b $}{H to bb}}

In this section we will obtain our final values for the higher-order coefficients of the perturbative expansion of $\operatorname{Im}\Pi(s)$.
The final results will be based on the approximants of Tab.~\ref{tab:coefimpi}.
We will not use approximants that post-dict the $\mathcal{O}(\alpha_s^3)$ coefficient since in the \Lb limit we found that with only three coefficients the rational approximants lack information to correctly predict the series beyond the fourth or fifth order.
Approximants based on $\operatorname{Im}\Pi(s)$ give results that differ significantly from the other approximants, especially for orders $\alpha_s^6$ and higher. As discussed before, the general structure of the Borel transform of $\operatorname{\Im}\Pi(s)$ suggests that it is not optimal to work with this quantity. However, to remain maximally conservative, and bearing in mind that our primary interest is on the series for $\operatorname{\Im}\Pi(s)$, we keep these results in our final analysis, which lead to larger (but very conservative) errors.

We start by computing the independent coefficients of the perturbative expansion of $\Pi(s)$, $d_{n,1}$, as we did in Sec.~\ref{sec:dn1largebeta0}. The
relation between the coefficients of $\Pi(s)$, $\operatorname{Im}\Pi(s)$, and $\Pi''(s)$ can be easily found from the expressions of
Sec.~\ref{sec:ScalarCorr} (additional useful formulas can be found in Ref.~\cite{Jamin:2016ihy}). In order to extract $d_{n,1}$ with $5\leq n\leq 8$
from our results given in Tab.~~\ref{tab:coefimpi}
we need, in principle, the coefficients of the
$\beta$ and $\gamma_m$ functions up to $\beta_7$ and $\gamma_m^{(8)}$, respectively. Since we know exactly only the coefficients up to five loops, we will
consider the unknown higher-order terms of the $\beta$ and $\gamma_m$ functions equal to zero, i.e., $\beta_6 = \beta_7 = 0$ and $\gamma_m^{(6)} = \gamma_m^{(7)} =
\gamma_m^{(8)} = 0$. This is a reasonable approximation since there is no sign of a possible divergence for these expansions
\cite{Baikov:2016tgj,Baikov:2014qja,Herzog:2017ohr}, in agreement with the (unproven) conjecture that the $\MSb$ scheme is a regular scheme, i.e., a
scheme where the $\beta$ and $\gamma_m$ functions are convergent series or at least do not diverge as fast as a factorial\cite{Beneke:1998ui}. As a check
of the reliability of this approximation, we also computed the coefficients $d_{5,1}$ zeroing the last known coefficients, $\beta_5$ and $\gamma_m^{(5)}$,
and compared with the results found using the known values of $\gamma_m^{(5)}$ and $\beta_5$. The difference did not exceed 0.21\%, which confirms that the
truncation of the $\beta$ and $\gamma_m$ function at the fifth term is, very likely, a very good approximation for our purposes. This assumption will be
used in the rest of this work.\footnote{As a further check of this assumption we have performed an estimate of the $\beta_6$ and $\gamma_m^{(6)}$ from PAs
built to their $\alpha_s$ expansion. Using these results, the shift we find in $d_{5,1}$ is of mere $\sim 0.34\%$ which is more than 10 times smaller
than the intrinsic uncertainty we find in $d_{5,1}$ from the PAs.}

\begin{table}[!t]
\centering
\caption{Final values for the perturbative coefficients of $\Pi(s)$, $d_{n,1}$, from the Pad\'e and D-log Pad\'e approximants to the perturbative expansions of $\operatorname{Im} \Pi(s)$ and $\Pi''(s)$ in QCD ($\MSb$ scheme, $N_f=5$).}
\begin{tabular}{cccc}
\thickhline
$d_{5,1}$ & $d_{6,1}$ & $d_{7,1}$ & $d_{8,1}$ \\ \hline
$(4.22 \pm 0.14) \times 10^4$ & $(5.79 \pm 0.35) \times 10^5$ & $(8.87 \pm 0.61) \times 10^6$ & $(1.51 \pm 0.11) \times 10^8$ \\ \thickhline
\end{tabular} \label{tab:dn1qcdfinal}
\end{table}

The coefficients $d_{n,1}$ were calculated from the estimated values of $c_n$ and $r_n$, the coefficients of the imaginary part and the second derivative of the scalar correlator respectively, given in Tab.~\ref{tab:coefimpi};
the final results for $d_{n,1}$ are in Tab.~\ref{tab:dn1qcdfinal}.\footnote{Performing our analysis with $N_f=3$ we find $d_{5,1} =77808 \pm 1400$ in excellent agreement with the recent estimate of Ref.~\cite{Jamin:2021qxb}, based on a model for the Borel transform.} The central values are calculated as the average between the largest and the smallest estimated coefficients. We assign an error to each coefficient that represents the maximum spread found between results from two approximants divided by two (a prescription that will be used through this work, and that is corroborated by explorations in \Lb).

With the same prescription we can also calculate the higher-order coefficients $c_n$ of the imaginary part of $\Pi(s)$, directly related to $\Gamma(H\to b \bar b)$. Our final result for the six-loop coefficient, $c_5$, the first unknown in QCD, is then
\beq\label{eq:c5final}
c_5 = -6900 \pm 1400 ,
\eeq
where the uncertainty is obtained from the spread in values from the different approximants, as explained above. Results up to $c_8$ are shown in Tab.~\ref{tab:cnqcd}. An important, if obvious, remark is that our errors should not be interpreted in a statistical sense. Rather, they give an interval where we expect the true value of the coefficients to lie.
Our final estimate for the intrinsic error in $c_5$, of about $20\%$, has a small impact in the sum of the perturbative series due to the suppression by $\alpha_s^5$--- as we will show in detail below. For the coefficients of sixth-order or higher, the errors associated are greater than 100\%, but they again do not lead to very large errors in the perturbative expansion. We remark that the estimated coefficients $c_n$ are not systematically sign-alternating, which suggests a competition between IR and UV renormalons at intermediate orders in QCD, in contrast with the typical situation in \Lb, as observed in related computations~\cite{Boito:2018rwt,Boito:2021wbj}.

Let us compare our result for $c_5$ with other estimates in the literature. This comparison is not completely
straightforward since other estimates do not have associated errors. The first
method, applied by Bakulev, Mikhailov and Stefanis \cite{Bakulev:2010gm}, models the coefficients of the series with two parameters, which are
determined through the known coefficients. With their estimated value for $d_{5,1}$ we
can calculate their central value for $c_5$, which is $-4052$; this result is not compatible with ours given the size and nature of our uncertainties in Eq.~(\ref{eq:c5final}). In Ref.~\cite{Bakulev:2010gm}
the coefficient $d_{5,1}$ is also calculated using the strategy employed by Kataev
and Starshenko \cite{Kataev:1995vh}, the Principle of Minimal Sensitivity (PMS), and the value obtained for $c_5$ is $-6886$ which is fully compatible
with our prediction.

It is also interesting to extract a final estimate for the Borel integral of the reduced $\operatorname{Im}\Pi(s)$, the function $F(a_s)$ of Eq.~(\ref{eq:impiqcdvalor}), which corresponds to an
estimate of the all-order true value of the series. In order to obtain this value, we calculated the perturbative series of $\operatorname{Im} \Pi(s)$
predicted by each approximant of Tab.~\ref{tab:coefimpi}. The ambiguity of the Borel integral,
associated with non-perturbative corrections and quantified by its imaginary part, is tiny in the application to Higgs decays, where the typical
scale is $m_H$. We have checked that for all practical purposes it can be neglected. Therefore, in this case, since the integrand is suppressed exponentially, the representative value of the integral
can be obtained by simply integrating the Taylor expansion of the Borel transformed $\operatorname{Im}\Pi(s)$. We have checked the reliability of this procedure in
cases where an analytical integration of the PAs was possible, and found that it leads to stable and correct results.\footnote{Another way of obtaining the
representative value of the Borel integral is to build higher-order PAs to the Taylor expansion of the relevant Borel transform. This procedure has
been used as an additional cross-check of our results.}
For $\alpha_s(m_H) = 0.1125 \pm 0.0009$, the final value obtained for the reduced $\operatorname{Im}\Pi(s)$ is
\beq
F = 0.2405 \pm (0.0022)_{\alpha_s} \pm (0.0002)_{\mathrm{PA}}, \label{eq:borelimpiqcd}
\eeq
where the first error is due to the uncertainty in the strong coupling, which largely dominates, and the second is due to the spread in the results from different approximants.
Our result is in good agreement with the one determined through the Principle of Maximum Conformality (PMC) \cite{Du:2018dma,Wu:2019mky}, which yields $0.2405 \pm 0.0001$, where the uncertainty is intrinsic to the method.

\begin{table}[!t]
\centering
\caption{Final values for the QCD perturbative coefficients $c_n$ of $\operatorname{Im}\Pi(s)$ ($\MSb$ scheme, $N_f=5$).}
\begin{tabular}{cccc}
\thickhline
$c_5$ & $c_6$ & $c_7$ & $c_8$ \\ \hline
$-6900 \pm 1400$ & $(0.3 \pm 3.5) \times 10^4$ & $(3.7 \pm 2.5) \times 10^5$ & $(0.2 \pm 2.4) \times 10^6$ \\ \thickhline
\end{tabular} \label{tab:cnqcd}
\end{table}

\begin{figure}[!t]
\centering
\includegraphics[width=0.7\textwidth, trim = 0cm 0cm 0cm 0cm, clip]{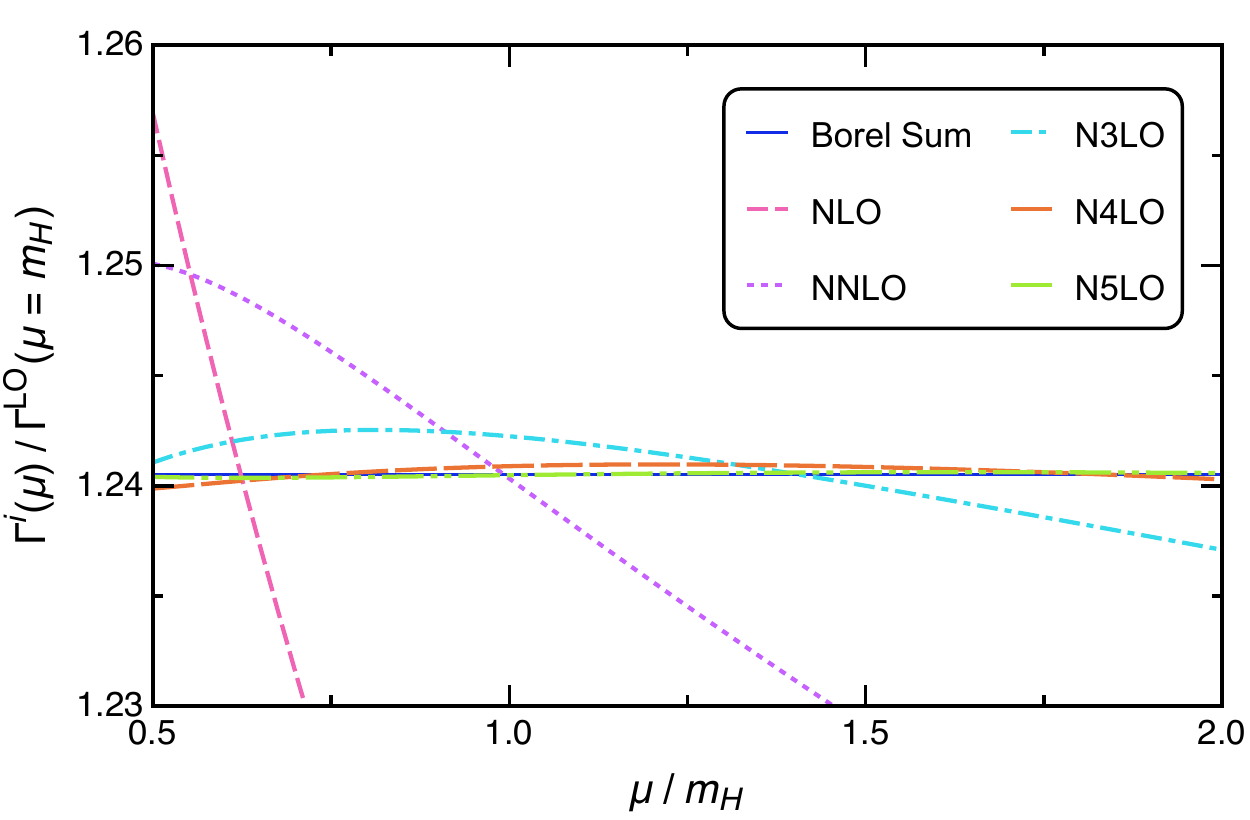}
\caption{Renormalization scale dependence of the normalized decay width given in Eq.~(\ref{eq:RenormDep}).}
\label{fig:decayrate}
\end{figure}

We apply now our final results to an analysis of the uncertainties in the SM calculation of $\Gamma(H \rightarrow b\bar{b})$. Let us start from a
discussion of the residual renormalization scale dependence order by order. As customary in the literature~\cite{Mondini:2019gid}, we study the ratio
\beq\label{eq:RenormDep}
\frac{\Gamma(H\to b\bar b)(\mu)}{\Gamma(H\to b\bar b)_{\rm LO}(m_H)} = \frac{m_b^2(\mu)}{m_b^2(m_H)}\sum_{n=0} a_s^n(\mu)\sum_{j=0}^n c_{n,j} \ln^j\left(m_H^2/\mu^2\right),
\eeq
where the dependent coefficients $c_{n,j}$ can be obtained using the RGE and $c_{n,0}\equiv c_n$. We calculate this quantity as a function of the
renormalization scale $\mu$, which was varied in the range\footnote{The values for the running
coupling and the running bottom-quark mass were obtained using our own code and with RunDec~\cite{Chetyrkin:2000yt,Herren:2017osy}, with perfect
agreement between the two.} $m_H/2\leq \mu \leq 2m_H$, and the final result up to N5LO,
computed using our prediction for the $c_5$ value of Eq.~(\ref{eq:c5final}), is in Fig.~\ref{fig:decayrate}. At N4LO the renormalization-scale is already mild and it is further reduced
at N5LO calculated from our value of $c_5$, as expected.

\begin{figure}[!t]
\begin{center}
\subfigure[]{\includegraphics[width=0.49\textwidth]{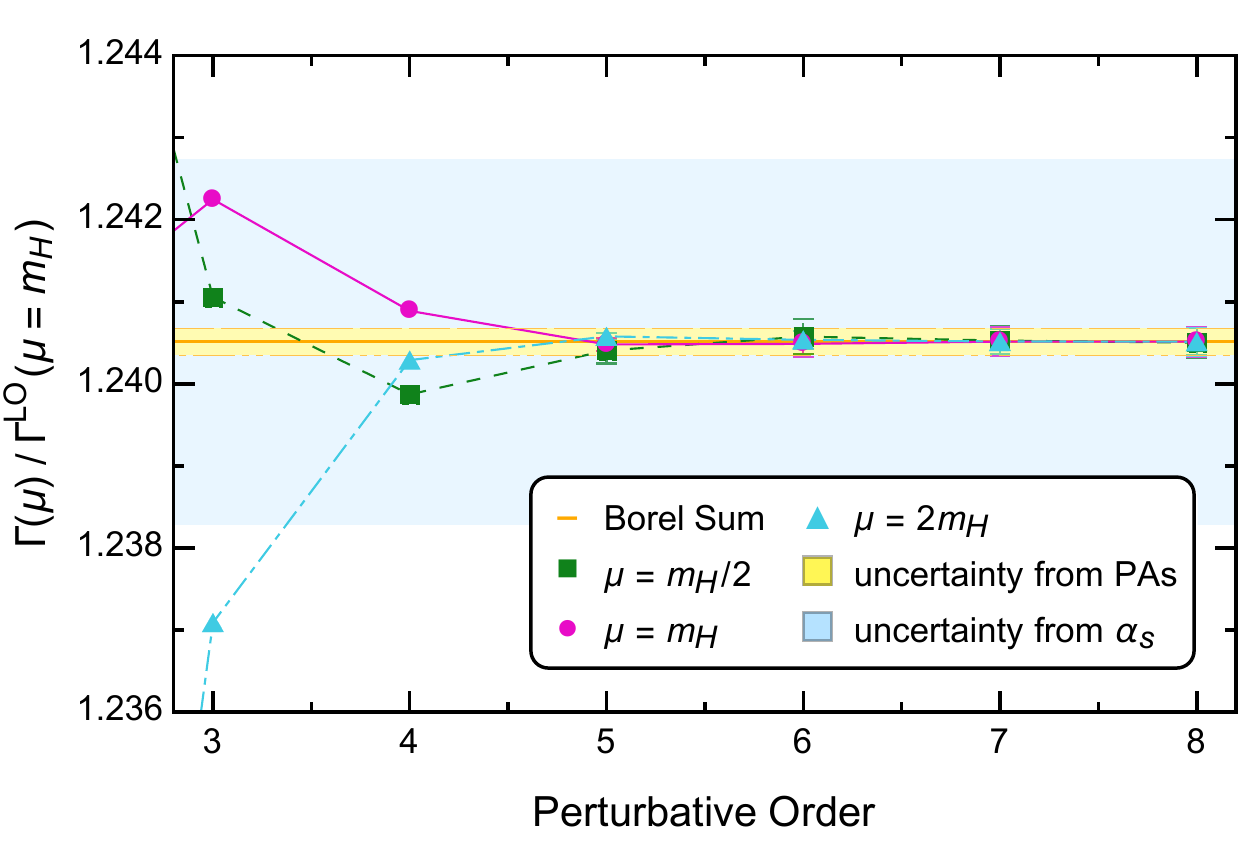}\label{fig:impiprevdifmu}}
\subfigure[]{\includegraphics[width=0.49\textwidth]{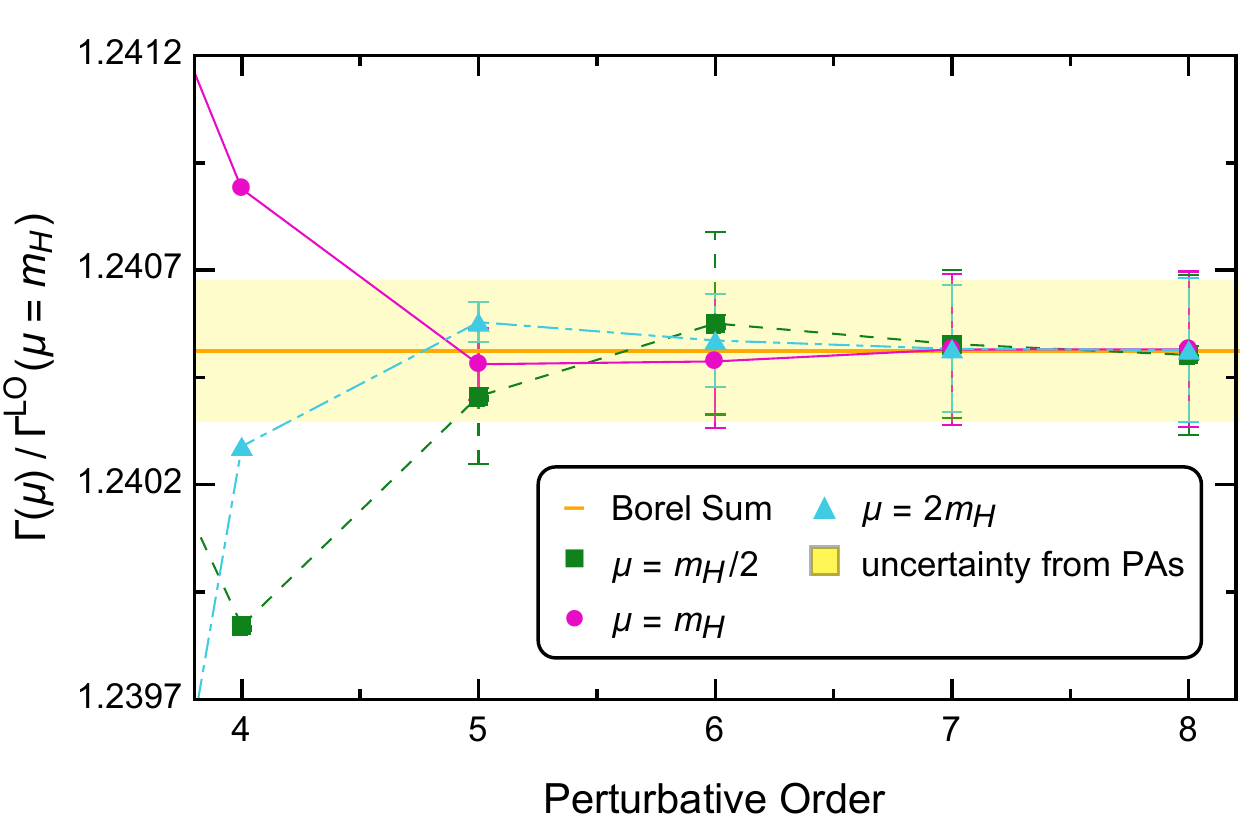}\label{fig:impiprevdifmu2}}
\caption{Perturbative expansion of the normalized decay rate Eq.~(\ref{eq:RenormDep}) in QCD at three different renormalization scales using the coefficients of Tab.~\ref{tab:cnqcd} for the coefficients $c_{n\geq 5}$ and the integral given in \cref{eq:borelimpiqcd}. The uncertainty of the Borel integral is due to (a) the strong coupling uncertainty (light-blue band) and (b) the different Pad\'e predictions (light-yellow band). The uncertainty from $m_b$ essentially cancels in Eq.~(\ref{eq:RenormDep}). The error bars of the points in (b) are due to the errors of the predicted coefficients. The $\alpha_s$ uncertainty is omitted in (b) since it would be too large for the scale of the figure.}\label{fig:seriesQCD}
\end{center}
\end{figure}

Numerically, the perturbative evaluation of $\Gamma(H\to b\bar b)(\mu)$ with $\mu=m_H$ up to our predicted contribution at N5LO, order by order, gives
\beq
\frac{\Gamma(H\to b\bar b)(m_H)}{\Gamma(H\to b\bar b)_{\rm LO}(m_H)} = 1 + 0.20295 + 0.03738 + 0.00192 - 0.00136 - \boxed{0.00041(8)_{\rm PAs}},\label{eq:decaycorrections}
\eeq
where the boxed term is the predicted N5LO result with the uncertainty stemming from the approximants. We display this series, order by
order in perturbation theory, for three different choices of the renormalization scale $\mu$, in
Figs.~\ref{fig:impiprevdifmu}~and~\ref{fig:impiprevdifmu2}. The error bars give the error from the series coefficients in Tab.~\ref{tab:cnqcd}. The
horizontal bands show the predicted value for the all order result, Eq.~(\ref{eq:borelimpiqcd}), with the uncertainties from $\alpha_s$ and from
the PAs. We see that at N4LO, due to the reduced renormalization scale dependence, the final uncertainty starts to be dominated by
$\alpha_s$. At N5LO and beyond, the series behaves essentially as if it had already converged with a tiny $\mu$-dependence and an excellent
agreement with our predicted all-order result. Our results for the 6-loop coefficient and for the estimated true value of the series confirm that the QCD perturbative series is under excellent control for this observable
although the precision in $\alpha_s$ (and $m_b$) must be increased in order to make the most of the perturbative calculation.

At N5LO, we find for the decay width of the Higgs into bottom quarks
\begin{align}
\Gamma(H \rightarrow b\bar{b}) = 2.3806 & (^{+0.041}_{-0.027})_{m_b} \pm (0.0042)_{\alpha_s} \nn \\
&\pm (0.0032)_{m_H} \pm(0.0002)_{\mu} \pm (0.0003)_{\mathrm{PAs}} \,\, \mathrm{MeV},
\end{align}
where the uncertainty marked with $\mu$ refers to renormalization scale variation.\footnote{Our central value agrees well with other estimates found in the literature~\cite{Stefanis:2009kv,Wang:2013bla}.} In this result we have used $m_b(m_b)=4.18^{+0.03}_{-0.02}$~GeV, $m_H=125.25\pm0.17$~GeV, $\alpha_s(m_Z)=0.1179\pm0.0010$~\cite{Zyla:2020zbs}, together
with our result $c_5=-6900\pm1400$. The use of the all-order estimate of Eq.~(\ref{eq:borelimpiqcd}) would lead to an almost identical result, since at
N5LO the series is reaching its true value. The uncertainty from renormalization scale variation was calculated as half of the maximum dispersion of the decay rate found when the scale was varied in the interval $m_H/2 \leq\mu\leq 2 \, m_H$. The inclusion of our N5LO result reduces the error due to scale variation by a factor of $4.3$ (it would be $\pm 0.001$ at N4LO).\footnote{We are tacitly assuming that $m_b$ and $\alpha_s$ are renormalized at the same scale $\mu=\mu_\alpha=\mu_m$. We could consider an independent scale variation~\cite{Dehnadi:2015fra,Boito:2020lyp}, which is more conservative and would lead to larger errors, but we do not expect any significant change in our conclusions from such a procedure.} The largest contributions to the error arise from the QCD parameters, $m_b$ and $\alpha_s$, as
well as the Higgs mass. (The size of our uncertainties from these parameters agrees with those of~\cite{Freitas:2019bre} when the same input values are used.) This is an example of a process where perturbative QCD is under excellent control, as could be inferred from the perturbative uncertainties associated with the result at $\mathcal{O}(\alpha_s^4)$, and the intrinsic uncertainty from the truncation
of the series is tamed for the present purposes. With our estimated N5LO result, and conservatively adding in quadrature the errors from scale variation and from the PAs, the truncation uncertainty does not exceed $0.02\%$.

\section{Conclusions}
\label{sec:conclusions}

We have applied the Pad\'e-Borel method to study missing higher orders (MHOs) in the massless scalar-current quark correlator. The method we use
was first applied to the Adler function in Ref.~\cite{Boito:2018rwt}. We make use of the knowledge available in the large-$\beta_0$ limit
to guide our study in QCD. This is important given that the available information about the series expansion in QCD is not abundant: only the first four non-trivial terms are known. In particular, the results in \Lb are used in order to select the variants of the approximants that lead to faster convergence with only four coefficients used as input. They are also instrumental for the QCD analysis, since the general structure of the Borel transforms in QCD can be inferred from the \Lb results.

Our main result is the prediction for the MHOs in $\operatorname{Im} \Pi(s)$, which is directly connected to $\Gamma(H\to b\bar b)$. We forecast the six-loop result to be $c_5=-6900\pm 1400$. We have shown that with this result the series is essentially immune to renormalization scale variations and the perturbative uncertainty becomes tiny: its does not exceed $0.02\%$. Our predictions for the MHOs and for the true value of the series indicate that the perturbative expansion is very well behaved even at higher orders and approaches smoothly the true value as predicted by the rational approximants, as can be seen in Fig.~\ref{fig:seriesQCD}. Although this could be inferred from the analysis of the series truncated at $\mathcal{O}(\alpha_s^4)$, it is reassuring to see it confirmed after the inclusion of our estimates for higher orders. Additionally, the coefficients of Tab.~\ref{tab:coefimpi} do not show a systematic sign alternation, which implies that the dominance of the UV renormalon is still not established.

As far as the SM uncertainty in $\Gamma(H\to b\bar b)$ is concerned, higher-loop calculations for the contributions discussed here, namely those associated with the massless scalar correlator, are probably not warranted. For the purposes of
matching the experimental uncertainty that should be achieved in the FCC-ee~\cite{Heinemeyer:2021rgq,Freitas:2019bre}, for example, the result at 5 loops together with our estimate of the MHOs should
suffice. The SM precision will be driven by the progress that can be made in the determination of $m_b$ and, to a lesser extend, of $\alpha_s$.

\section*{Acknowledgements}
We thank the anonymous referee for valuable comments on a previous version of this manuscript.
DB thanks the University of Vienna and the Universitat Aut\`onoma de Barcelona, where part of this work was carried out, for hospitality. DB's work was supported in part by the S\~ao Paulo Research Foundation (FAPESP) Grant
No.\ 2015/20689-9, and by CNPq Grant No.\ 309847/2018-4. The work of CYL was financed in part by FAPESP grants No.\ 2018/21050-0 and No.\ 2020/15532-1.
DB and CYL received partial support from Coordenação de Aperfeiçoamento de Pessoal de Nível Superior – Brasil (CAPES) – Finance Code 001.
The work of PM was supported by the Spanish Ministry of Science and Innovation (PID2020-112965GB-I00/AEI/ 10.13039/501100011033) and from the
Agency for Management of University and Research Grants of the Government of Catalonia (project
SGR 1069).

\appendix
\section[QCD \boldmath \texorpdfstring{$\beta$}{beta} and \texorpdfstring{$\gamma_m$}{gammam} functions and scale invariant quark mass]{\boldmath QCD \boldmath \texorpdfstring{$\beta$}{beta} and \texorpdfstring{$\gamma_m$}{gammam} functions and scale invariant quark mass}
\label{app:betagamma}

Our definitions for the $\beta$ and $\gamma_m$ functions are
\begin{align}
&\beta(a_s) \equiv -\mu \, \dfrac{\mathrm{d} a_s}{\mathrm{d} \mu} = \beta_1 a_s^2 + \beta_2 a_s^3 + \beta_3 a_s^4 + \cdots , \\
&\gamma_m(a_s) \equiv -\frac{\mu}{m_q} \, \dfrac{\mathrm{d} m_q}{\mathrm{d} \mu} = \gamma_m^{(1)} a_s + \gamma_m^{(2)} a_s^2 + \gamma_m^{(3)} a_s^3 + \cdots ,
\end{align} with $a_s\equiv\alpha_s/\pi$.
For definiteness, we give the one-loop coefficients of these functions:
\begin{align}
\beta_1 = \frac{11}{2} -\frac{N_f}{3}, \qquad {\rm and} \qquad
\gamma_m^{(1)}= 2.
\end{align}
With these definitions, the RGI quark mass can be written as
\beq
m_q(\mu) \equiv \wh{m}_q \, [\alpha_s(\mu)]^{\gamma_m^{(1)}/\beta_1} \, \mathrm{exp} \left\{ \int\displaylimits_0^{a_s(\mu)} \mathrm{d}a \left[ \dfrac{\gamma_m(a)}{\beta(a)} - \dfrac{\gamma_m^{(1)}}{\beta_1 a} \right] \right\} . \label{eq:massrgi}
\eeq

\bibliographystyle{jhep}
\bibliography{References}

\end{document}